\documentclass[twocolumn,twoside,slac_two]{revtex4}
\usepackage{graphicx}
\usepackage{fancyhdr}
\usepackage{color}
\usepackage{amssymb}
\usepackage{amsmath}

\def\etal{{\it et al.}}
\def\ie{{\it i.e.}}
\newcommand{\newln}{\\&\quad\quad{}}

\input babarsym
\newcommand{\gevcccc}{\ensuremath{{\mathrm{\,Ge\kern -0.1em V^2\!/}c^4}}\xspace}

\def\Kmaybestar {\ensuremath{K^{(*)}\xspace}}
\def\afb {\mbox{${\cal A}_{FB}$}\xspace}

\def\ctk {\ensuremath{\cos\theta_{K}}\xspace}

\def\ctl {\ensuremath{\cos\theta_{\ell}}\xspace}

\def\fl {\mbox{$F_L$}\xspace}

\def\kll {\B\to\Kmaybestar\ellell\xspace}
\def\kmaybeee {\B\to\Kmaybestar\epem\xspace}

\def\kmaybemm {\B\to\Kmaybestar\mumu\xspace}

\def\mkpi {\ensuremath{m_{\kaon\pi}}\xspace}
\def\mll {\ensuremath{m_{\ell\ell}}\xspace}

\def\modekavgll {\ensuremath{B\to K\ellell}\xspace}

\def\modekstll {\ensuremath{B\rightarrow K^{*}\ellell}\xspace}

\begin{document}

\title{Recent $B\to X_s\gamma$ and $B\to X\ell\ell$ Results from $B$ Factories}


\author{L. Sun (on behalf of the \babar\, and Belle collaborations)}
\affiliation{University of Cincinnati, Cincinnati OH 45221, USA}
%
%\author{P. Lucas}
%\affiliation{FNAL, Batavia, IL 60510, USA}

\begin{abstract}
In this talk, a wide range of recently published results on rare $B$ decays from  \babar\, and Belle are 
covered. The decays of $B\to X_s\gamma$, $B\to \Kmaybestar\ellell$, and
lepton-number violation in $B$ decays are measured for new physics searches.
\end{abstract}

\maketitle

\thispagestyle{fancy}


\section{Introduction}
By 2008 and 2010, $\babar\,$ and Belle completed data-taking respectively. 
Both $B$ factories produced $\BB$ pairs at the $\FourS$ resonance with 
asymmetric-energy $e^+ e^-$ colliders. These $\BB$ pairs were collected with
$\babar\,$ and Belle detectors. Both detectors share similar designs  
and provide good discrimination on charged particles, especially for
$K^{\pm}$/$\pi^{\pm}$. Muons are also differentiated from charged hadrons.
High-energy photons and electrons are precisely measured
with electro-magnetic calorimeters in both detectors.

The \babar\, and Belle collaborations continue to produce interesting results on rare $B$ decays. 
Based on 471 million $\BB$ pairs from $\babar$\, this talk presents $b\to s \gamma$ transition rates and photon energy spectrum 
using a sum of exclusive modes, rates and asymmetries in
 exclusive $B\to \Kstar\ellell$ decays, and
 a search for lepton-number violation (LNV) processes in $B^+\to K^-\ell^+\ell^+$ decays\footnote{Charge conjugation is implied throughout except as explicitly
noted.}. In addition, this talk presents a search for LNV processes in $B^+\to D^-\ell^+\ell^+$ decays 
based on 773 million $\BB$ pairs from Belle.

\section{$B\to X_s \gamma$ Transition Rates and Photon Energy Spectrum with
a Sum of Exclusive Modes}

\begin{figure}
\includegraphics{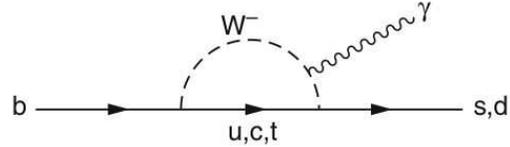}
\caption{\label{fig:btosgdiag}The leading order Feynman diagram for the transition $b\to (s/d)\gamma$ in the SM.~\cite{btosgbabar}}
\end{figure}

The $b\to s\gamma$ transitions are flavor-changing neutral-current (FCNC) processes
and forbidden at tree level in the Standard Model (SM). The leading-order 
radiative penguin diagram for this type of transition is 
shown in Figure~\ref{fig:btosgdiag}. 
For photon energy $E_\gamma > 1.6$~\gev, the SM-based prediction for
the decay rate is at ${\cal B} (\Bbar \to X_s\gamma) = (3.15\pm 0.23)\times 10^{-4}$~\cite{Misiak:2006zs}, with $X_s$ as 
the hadronic final state with strangeness. 
This prediction is in good agreement with the current world average of 
experimental results at ${\cal B} (\Bbar \to X_s\gamma) = (3.55\pm 0.25\pm0.09)\times 10^{-4}$ for the same photon energy 
cutoff at $E_\gamma > 1.6$~\gev~\cite{HFAG}. Here
the second uncertainty comes from the extrapolation of the photon energy 
shape function from the experimental photon energy to the 1.6~\gev cutoff. 
New physics beyond the SM may enter the radiative loop of $b\to s\gamma$ and alter the decay 
rate significantly. Therefore comparing the experimental results 
and SM-based predictions provides a good test of the SM. 
Furthermore, the photon energy spectrum is important for us to understand
the $b$ quark momentum distribution inside the $B$ meson. The shape
of this distribution is a critical input to the determination of the CKM matrix element \Vub
in inclusive charmless semileptonic B decays.
The measured energy spectrum can be fit to different models in order
to find the best values of $m_b$ and
$\mu^2_{\pi}$ as the heavy quark effective theory (HQET) parameters.

This $\babar\,$ analysis~\cite{btosgbabar} reconstructs the $X_s$ system in 38 different 
exclusive final states listed in 
Table~\ref{tab:xsmodes}. With the reconstructed $X_s$ system, 
    the energy of the transition photon in the $B$ rest frame $E_{\gamma}^B$ 
    is thus accessible through:
\begin{equation}
E_{\gamma}^B = \frac{m^2_B-m^2_{X_s}}{2m_B}\,,
\label{eq:Egammadef}
\end{equation}
where $m_B$ and $m_{X_s}$ are the invariant masses of
the $B$ meson and of the $X_s$ hadronic system, respectively.

\begin{table}%[htp]
\begin{small}
\begin{center}
\caption{\label{tab:xsmodes} The 38 final states for $\B \to X_s \gamma$ used in this analysis. Here
    $\KS$ is reconstructed through $\KS \to\pip\pim$.~\cite{btosgbabar}}
\begin{tabular}{cl|cl}
\hline 
\hline
Mode No.       & Final State                           &Mode No.  &Final State\\ 
\hline
1       & $ \KS\pi^{+}\gamma$             & 20        & $ \KS\pi^{+}\pi^{-}\pi^{+}\pi^{-}\gamma$\\
2       & $ K^{+}\pi^{0}\gamma$             & 21        & $ K^{+}\pi^{+}\pi^{-}\pi^{-}\pi^{0}\gamma$\\
3       & $ K^{+}\pi^{-}\gamma$             & 22        & $ \KS\pi^{+}\pi^{-}\pi^{0}\pi^{0}\gamma$\\
4       & $ \KS\pi^{0}\gamma$             & 23        & $ K^{+}\eta\gamma$\\
5       & $ K^{+}\pi^{+}\pi^{-}\gamma$          & 24        & $ \KS\eta\gamma$\\
6       & $ \KS\pi^{+}\pi^{0}\gamma$          & 25        & $ \KS\eta\pi^{+}\gamma$\\
7       & $ K^{+}\pi^{0}\pi^{0}\gamma$          & 26        & $ K^{+}\eta\pi^{0}\gamma$\\
8       & $ \KS\pi^{+}\pi^{-}\gamma$          & 27        & $ K^{+}\eta\pi^{-}\gamma$\\
9       & $ K^{+}\pi^{-}\pi^{0}\gamma$          & 28        & $ \KS\eta\pi^{0}\gamma$\\
10      & $ \KS\pi^{0}\pi^{0}\gamma$          & 29        & $ K^{+}\eta\pi^{+}\pi^{-}\gamma$\\
11      & $ \KS\pi^{+}\pi^{-}\pi^{+}\gamma$       & 30        & $ \KS\eta\pi^{+}\pi^{0}\gamma$\\
12      & $ K^{+}\pi^{+}\pi^{-}\pi^{0}\gamma$       & 31        & $ \KS\eta\pi^{+}\pi^{-}\gamma$\\
13      & $ \KS\pi^{+}\pi^{0}\pi^{0}\gamma$       & 32        & $ K^{+}\eta\pi^{-}\pi^{0}\gamma$\\
14      & $ K^{+}\pi^{+}\pi^{-}\pi^{-}\gamma$       & 33        & $ K^{+}K^{-}K^{+}\gamma$\\
15      & $ \KS\pi^{0}\pi^{+}\pi^{-}\gamma$       & 34        & $ K^{+}K^{-}\KS\gamma$\\
16      & $ K^{+}\pi^{-}\pi^{0}\pi^{0}\gamma$       & 35        & $ K^{+}K^{-}\KS\pi^{+}\gamma$\\
17      & $ K^{+}\pi^{+}\pi^{-}\pi^{+}\pi^{-}\gamma$    & 36        & $ K^{+}K^{-}K^{+}\pi^{0}\gamma$\\
18      & $ \KS\pi^{+}\pi^{-}\pi^{+}\pi^{0}\gamma$    & 37        & $ K^{+}K^{-}K^{+}\pi^{-}\gamma$\\
19      & $ K^{+}\pi^{+}\pi^{-}\pi^{0}\pi^{0}\gamma$    & 38        & $ K^{+}K^{-}\KS\pi^{0}\gamma$\\
\hline
\hline
\end{tabular}
\end{center}
\end{small}
\end{table}

\begin{figure}
\includegraphics[width=0.8\linewidth]{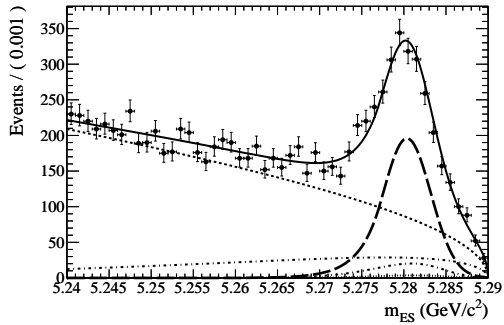}
\caption{\label{fig:bsgfit}\babar: $\mes$ fit example in the region 1.4 $< m_{X_s}<$ 1.5~\gevcc. Shown
are data (points with error bars) along with the total fit (solid curve), 
signal (thick dashed curve), and different background contributions (other curves, see Ref.~\cite{btosgbabar} for
        detailed description).}
\end{figure}

In each event, this analysis requires
at least one photon candidate with $1.6<E_{\gamma}^{*} < 3.0$~\gev in the center-of-mass (CM) frame. This analysis also requires
$m_{X_s}$ to be 
within a range of 0.6 and 2.8~\gevcc, and divides this range into 18 bins. 
These 18 $m_{X_s}$ bins include 14 bins each with a width of 100~\mevcc for $m_{X_s}<2.0$~\gevcc,
and 4 bins each with a width of 200~\mevcc for $m_{X_s}\geq 2.0$~\gevcc.
Common to all other \babar\, analyses,
the beam-energy substituted mass $\mes = \sqrt{E^{*2}_{\rm beam}-p^{*2}_B}$ is defined, where $E^*_{\rm beam}$ and $p^*_B$ are the beam energy and
the $B$ meson momentum in the CM frame, respectively.
In each $m_{X_s}$ bin, the $\mes$ distribution is fit to extract the signal 
yield. An example of the $\mes$ fit for $1.4<m_{X_s}<1.5$~\gevcc is shown in
Fig.~\ref{fig:bsgfit}.

\begin{figure}
\includegraphics[width=0.95\linewidth]{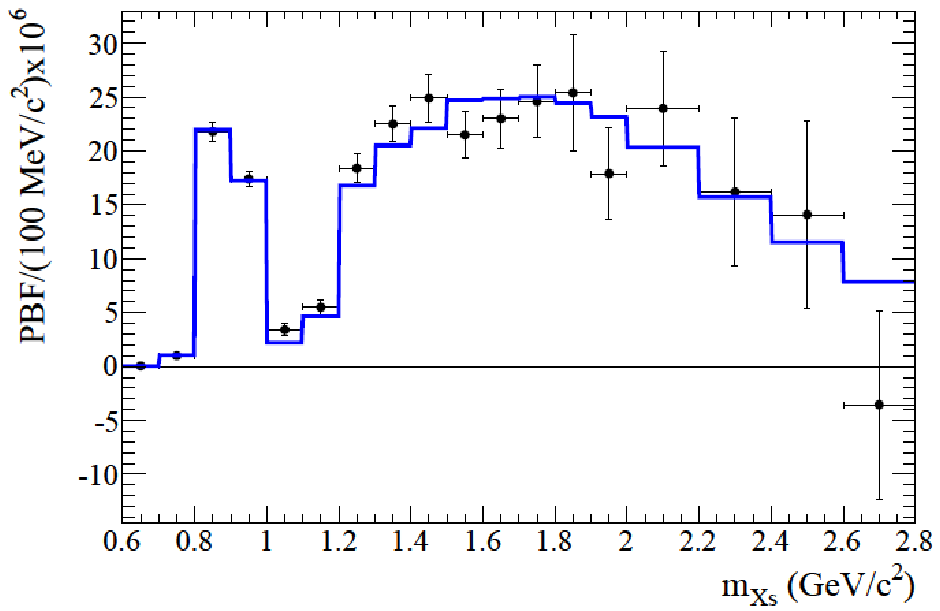}\\
\includegraphics[width=0.95\linewidth]{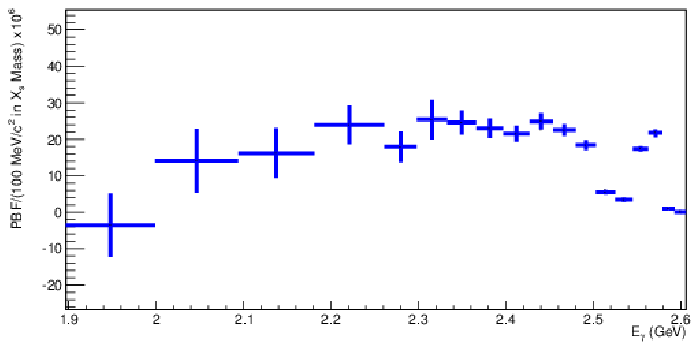}
\caption{\label{fig:xsmassspect}\babar: Top: Measured partial branching fractions (points with error bars) in the 18 $m_{X_s}$ bins, 
with the spectrum fit using the kinetic model (solid line). Bottom: Measured partial branching fractions (points with error bars) in the 
corresponding photon energy bins. All the error bars shown here include the statistical and systematic uncertainties added
in quadrature.~\cite{btosgbabar}}
\end{figure}

Figure~\ref{fig:xsmassspect} shows the measured partial branching
fractions in the 18 $m_{X_s}$ bins, as well as in the corresponding
$E_\gamma$ bins. The measured $m_{X_s}$ spectrum is fit with two different models: the ``kinetic model''~\cite{Benson:2004sg} and ``shape function model''~\cite{Lange:2005yw}, to find
the best values for
$m_b$ and $\mu^2_{\pi}$ as summarized in Table~\ref{tab:hqetfits}. 
These numbers are compatible with the world averages for these two models~\cite{HFAG}
also listed in Tables~\ref{tab:hqetfits}. Figure~\ref{fig:xsmassspect} also
shows the fit with the kinetic model as an example. 
By combining the partial branching fraction results, the total branching
fraction for $E_\gamma > 1.9$~\gev is ${\cal B}(\Bbar \to X_s \gamma) = (3.29\pm0.19\pm0.48)\times 10^{-4}$, where the first and second uncertainties
are statistical and systematic, respectively. 

\begin{table}%[htp]
\begin{center}
\caption{The best fit HQET parameter values based on the measured $m_{X_{s}}$ spectrum,
    are compared to the world averages (``HFAG'').~\cite{btosgbabar}\label{tab:hqetfits}}
\begin{tabular}{c|cc}
\hline
\hline
& Kinetic model & Shape function model \\ \hline
$m_{b}$ (\gevcc)            & $4.568^{+0.038}_{-0.036}$  &  $4.579^{+0.032}_{-0.029}$\\
HFAG:             & \textcolor{red}{$4.591\pm 0.031$}  &  \textcolor{red}{$4.620^{+0.039}_{-0.032}$} \\
\hline
$\mu_{\pi}^{2}$ ($\gev^{2}$) & $0.450^{+0.054}_{-0.054}$ & $0.257^{+0.034}_{-0.039}$ \\
HFAG:             & \textcolor{red}{$0.454\pm 0.038$}  &  \textcolor{red}{$0.288^{+0.054}_{-0.074}$} \\
\hline
\hline
\end{tabular}
\end{center}
\end{table}

\section{Exclusive $B\to \Kmaybestar \ellell$ Decays}
\begin{figure}%[b]
%\rule{5cm}{0.2mm}\hfill\rule{5cm}{0.2mm}
%\vskip 2.5cm
%\rule{5cm}{0.2mm}\hfill\rule{5cm}{0.2mm}
\begin{center}
\includegraphics[height=3cm]{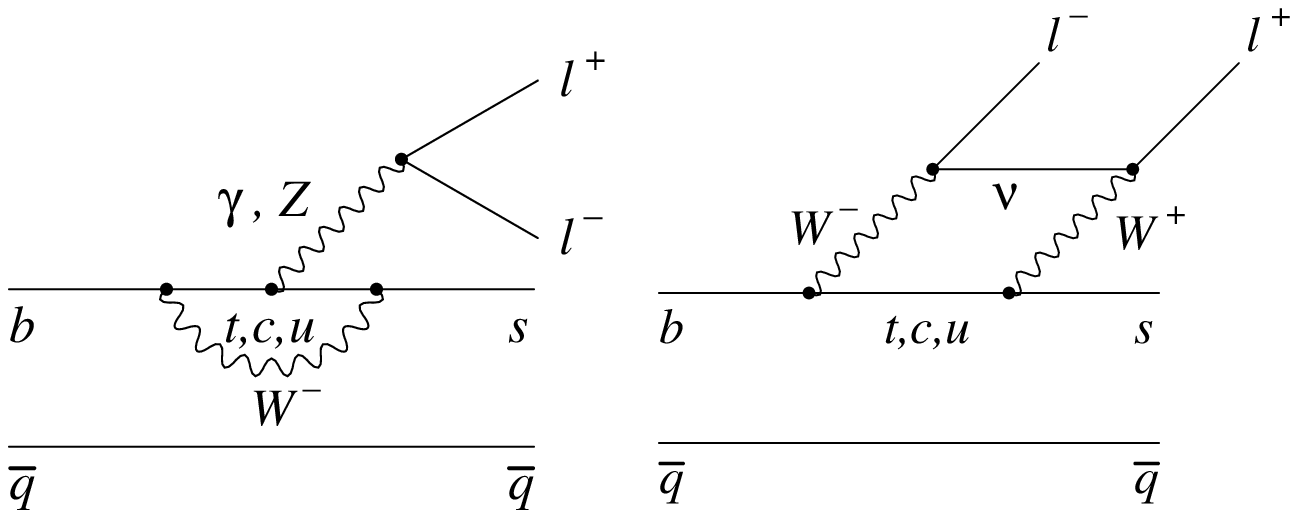}\\\vspace{5 mm}
\includegraphics[width=0.31\linewidth]{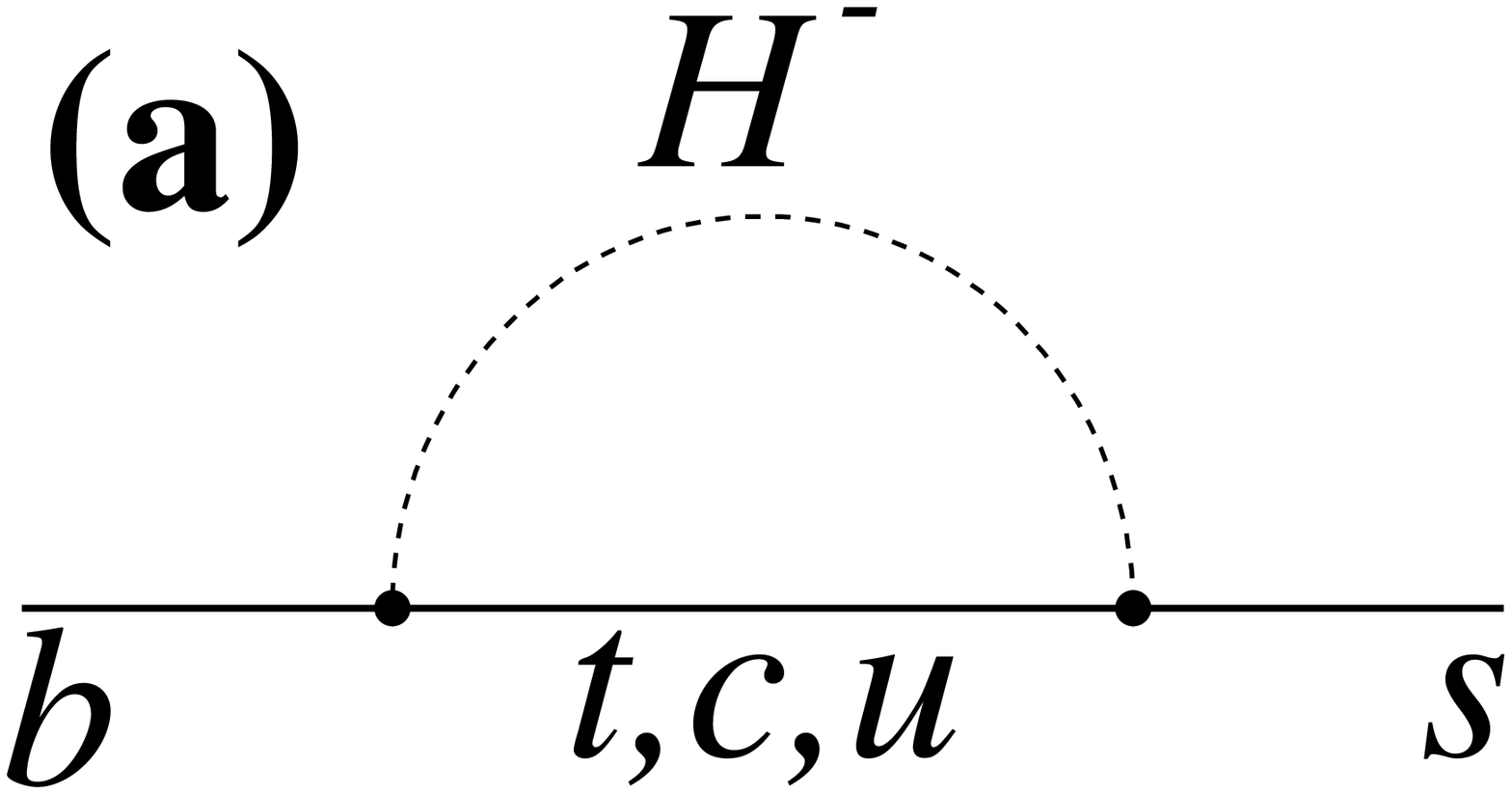}
\includegraphics[width=0.31\linewidth]{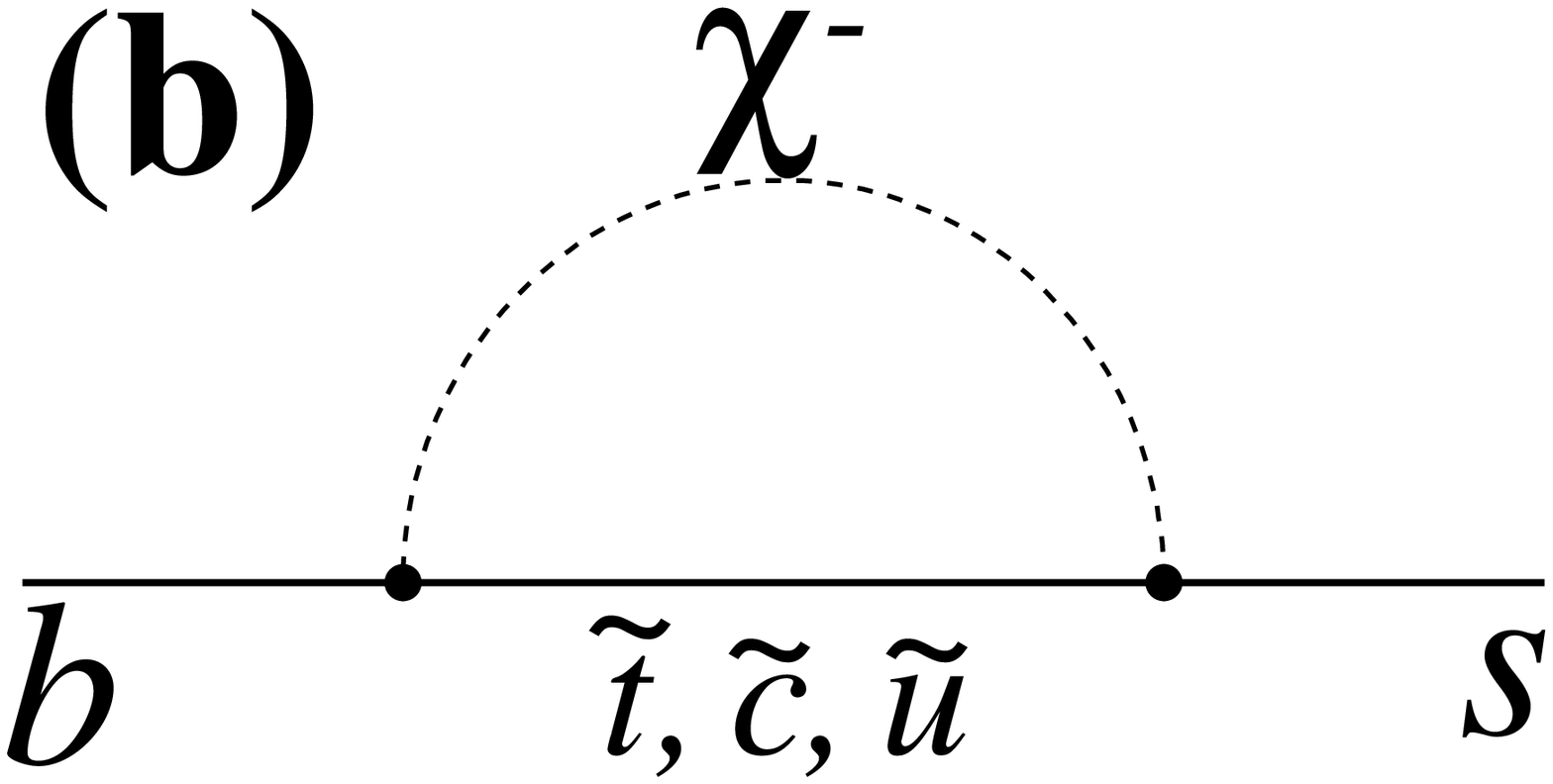}
\includegraphics[width=0.31\linewidth]{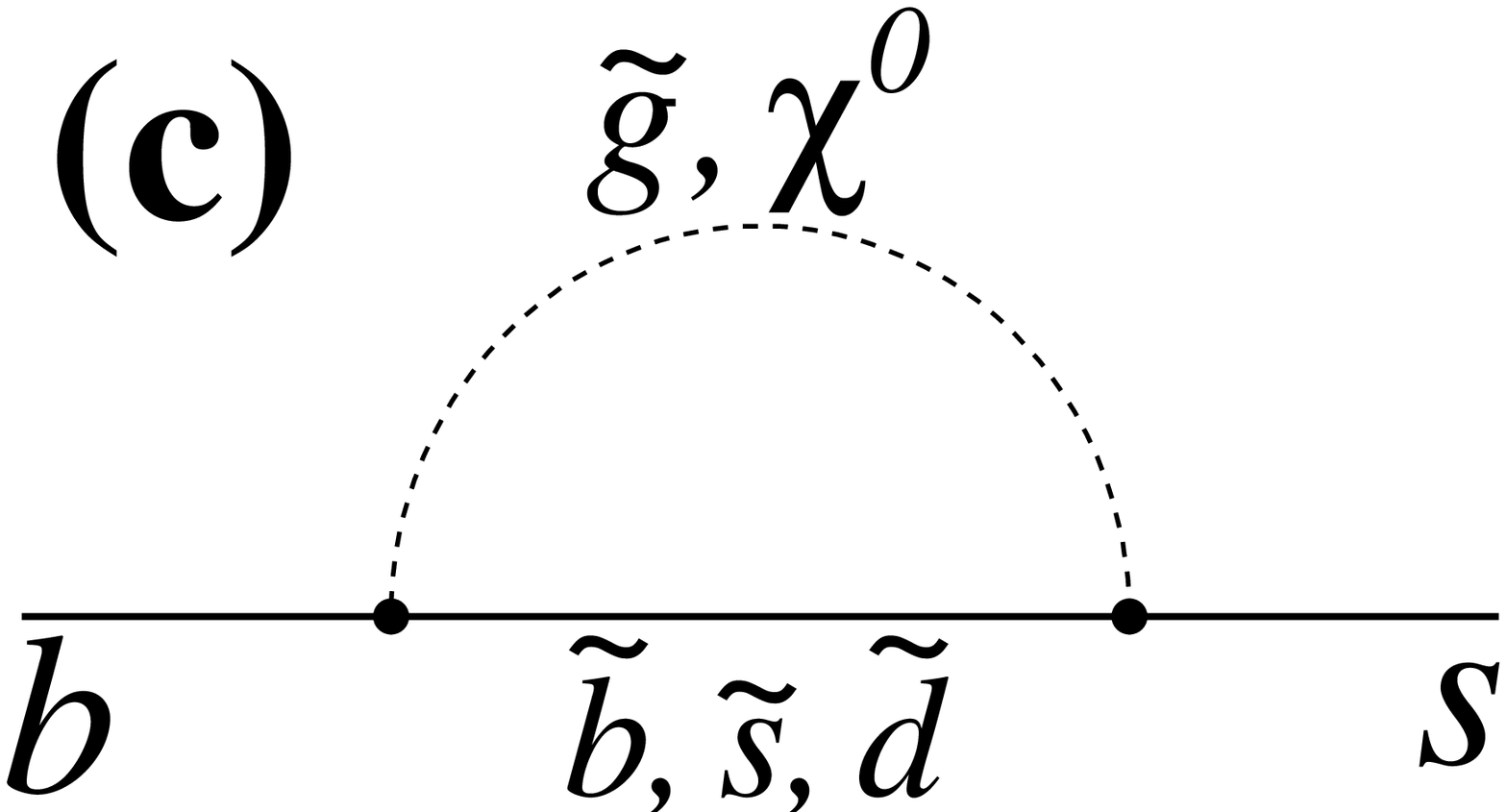}
\caption{Top: Lowest-order Feynman diagrams for $\btosll$ in the SM.
Bottom: Examples of new physics loop contributions to $\btosll$: (a) charged Higgs ($H^-$); (b) squark ($\tilde{t}, \tilde{c}, \tilde{u}$) and chargino ($\chi^-$); (c) squark ($\tilde{b}, \tilde{s}, \tilde{d}$) and gluino ($\tilde{g}$)/neutralino ($\chi^0$).~\cite{kllpaper}}
\label{fig:slldiag}
\end{center}
\end{figure}

Similar to $b\to s\gamma$, the FCNC $b\to s\ellell$ processes are also forbidden
at tree level in the SM. The $B\to \Kmaybestar \ellell$ decays are allowed in 
the loop and box diagrams as shown in Fig.~\ref{fig:slldiag} with 
branching fractions at about $10^{-6}$. The effective Hamiltonian 
for these decays factorizes short-distances contributions represented
by the Wilson coefficients from long-distance effects. Three effective 
Wilson coefficients are relevant here: $C_7^{\rm eff}$ from the electromagnetic 
penguin diagram, $C_{9}^{\rm eff}$ and $C_{10}^{\rm eff}$ 
from the vector part and the axial-vector part of the $Z$ penguin
and $W^+ W^-$ box diagrams, respectively~\cite{Buchalla}. 
New physics may bring in new loops involving
particles such as charged Higgs, squarks, neutralinos, and charginos
particles as depicted in
the bottom of Figure~\ref{fig:slldiag}. 
These loop contributions may be comparable to the SM ones, and
alter the effective 
Wilson coefficient values from their SM expectations~\cite{isospin}.

\begin{figure}%[b]
\begin{center}
\includegraphics[width=0.8\linewidth]{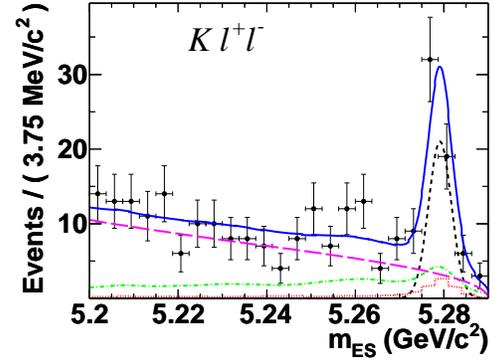}
\caption{\babar: Example \mes fit in the region $10.1\le s<12.9$~\gevcccc for $B\to K\ellell$.
Data (points with error bars) are shown along with the total fit (solid curve), signal (short-dashed
curve), and different background contributions (other curves, see Ref.~\cite{kllpaper} for
        detailed description).
}
\label{fig:kllfit}
\end{center}
\end{figure}

In this \babar\, analysis, \kll signal events are reconstructed in eight final states with
an $\epem$ or $\mumu$ pair, and a $\KS$, $\Kp$, $\Kstarp(\to\KS\pip)$,
   or $\Kstarz(\to\Kp\pim)$, where a $\KS$ candidate
   is reconstructed in the $\pipi$
  final state. Another final state $\Kp\piz\epem$ is included only for the
  $K^*\ellell$ angular measurements introduced in Sect.~\ref{sec:kstllang}.
  Selected $\Kstar$ candidates are also required to have an invariant
  mass of $0.72<\mkpi<1.10$~\gevcc.
The measurements are performed in six bins of di-lepton mass squared $s\equiv\mll^2$:
$0.1\le s<2.0$~\gevcccc, $2.0\le s<4.3$~\gevcccc, $4.3\le s<8.1$~\gevcccc,
$10.1\le s<12.9$~\gevcccc, $14.2\le s<16.0$~\gevcccc, and $ s\ge 16.0$~\gevcccc.
In the analysis, two $s$ regions are vetoed to minimize the $\jpsi$ and $\psitwos$ contributions.
The binning choices are largely consistent with those used in the Belle~\cite{belle09},
    CDF~\cite{cdf11}, and LHCb~\cite{lhcb11} experiments.
An example $\mes$ fit to extract signal $K\ellell$ events is depicted in Fig.~\ref{fig:kllfit}.
The experimental details are presented in Ref.~\cite{kllpaper}.

\subsection{Rates and Rate Asymmetries}
\begin{figure}%[b!]
\begin{center}
\includegraphics[height=4.0cm]{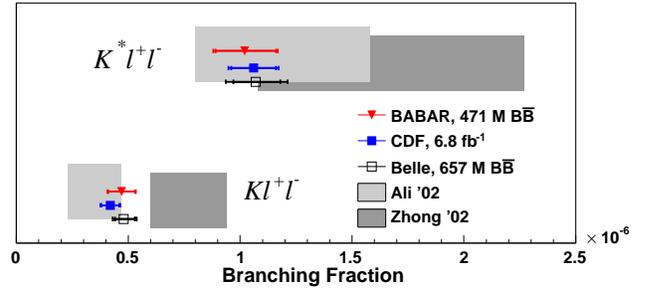}
\caption{\babar: Total branching fractions for 
    $K \ell^+ \ell^-$ and  $K^* \ell^+ \ell^-$ compared
       with Belle~\protect\cite{belle09} and CDF~\protect\cite{cdf11} 
       measurements and with 
    predictions from the Ali~\etal~\protect\cite{Ali:2002jg}, 
and Zhong~\etal~\protect\cite{Zhong:2002nu} models.~\cite{kllpaper} }
\label{fig:klltbf}
\end{center}
\end{figure}

\begin{figure}%[b!]
\begin{center}
\includegraphics[height=6.0cm]{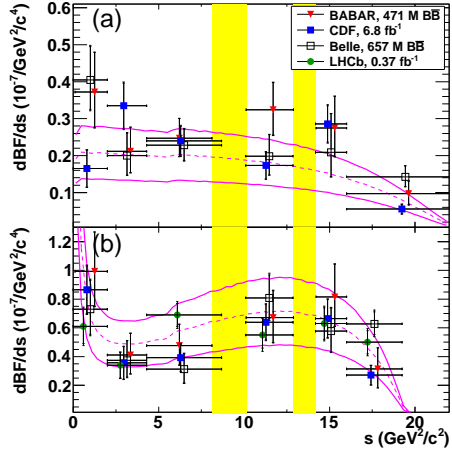}
\caption{\babar: Partial branching fractions for (a) $K \ell^+ \ell^-$ and (b) $K^* \ell^+ \ell^-$ as a function of $s$ showing \babar\, measurements, Belle measurements~\protect\cite{belle09}, CDF measurements~\protect\cite{cdf11}, LHCb measurements~\protect\cite{lhcb11}, and the SM prediction from the Ali~\etal~model~\protect\cite{Ali:2002jg} with $B\to\Kmaybestar$ form factors~\protect\cite{ffmodels} (magenta dashed lines).
    %The CDF results
%   are only based on final states $\Bp\to \Kp\mumu$ and $\Bz\to\Kstarz\mumu$,
%       for $K \ell^+ \ell^-$ and $\Kstar\ellell$ modes, respectively.
       The magenta solid lines show the theory uncertainties. The vertical 
       yellow shaded bands show the vetoed $s$ regions around the $J/\psi$ and $\psitwos$ resonances.~\cite{kllpaper} }
\label{fig:kllpbf}
\end{center}
\end{figure}

The measured total branching fractions are:
\begin{eqnarray}
{\cal B} (\modekavgll) & = & (4.7\pm0.6\pm 0.2) \times 10^{-7},\nonumber\\
{\cal B} (\modekstll) & = & (10.2_{-1.3}^{+1.4}\pm 0.5) \times 10^{-7},\nonumber
\end{eqnarray}
where the first and second uncertainties are statistical and systematic, respectively.
In Fig.~\ref{fig:klltbf}, the measured total branching fraction results agree well
with the results from  the Belle and CDF experiments~\cite{belle09,cdf11},
     as well as two sets of SM-based predictions~\cite{Ali:2002jg,Zhong:2002nu}.
Figure~\ref{fig:kllpbf} shows the \babar\, partial branching fraction results
in agreement with the Belle, CDF, and LHCb results~\cite{belle09,cdf11,lhcb11}. These \babar\, results are
also consistent with the SM-based predictions~\cite{Ali:2002jg,ffmodels}.

\begin{figure}%[b!]
\begin{center}
\includegraphics[height=6cm]{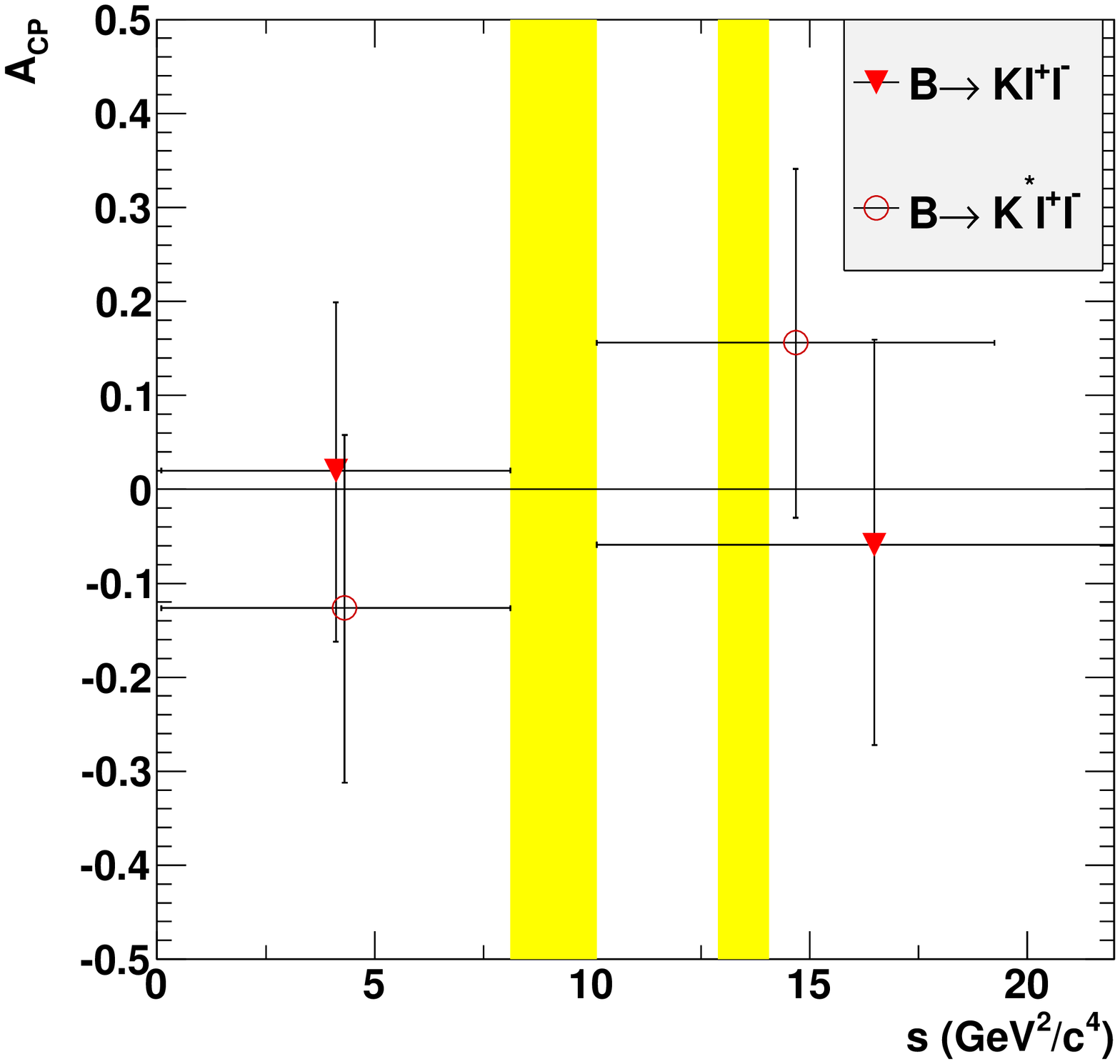}\\
\includegraphics[height=6cm]{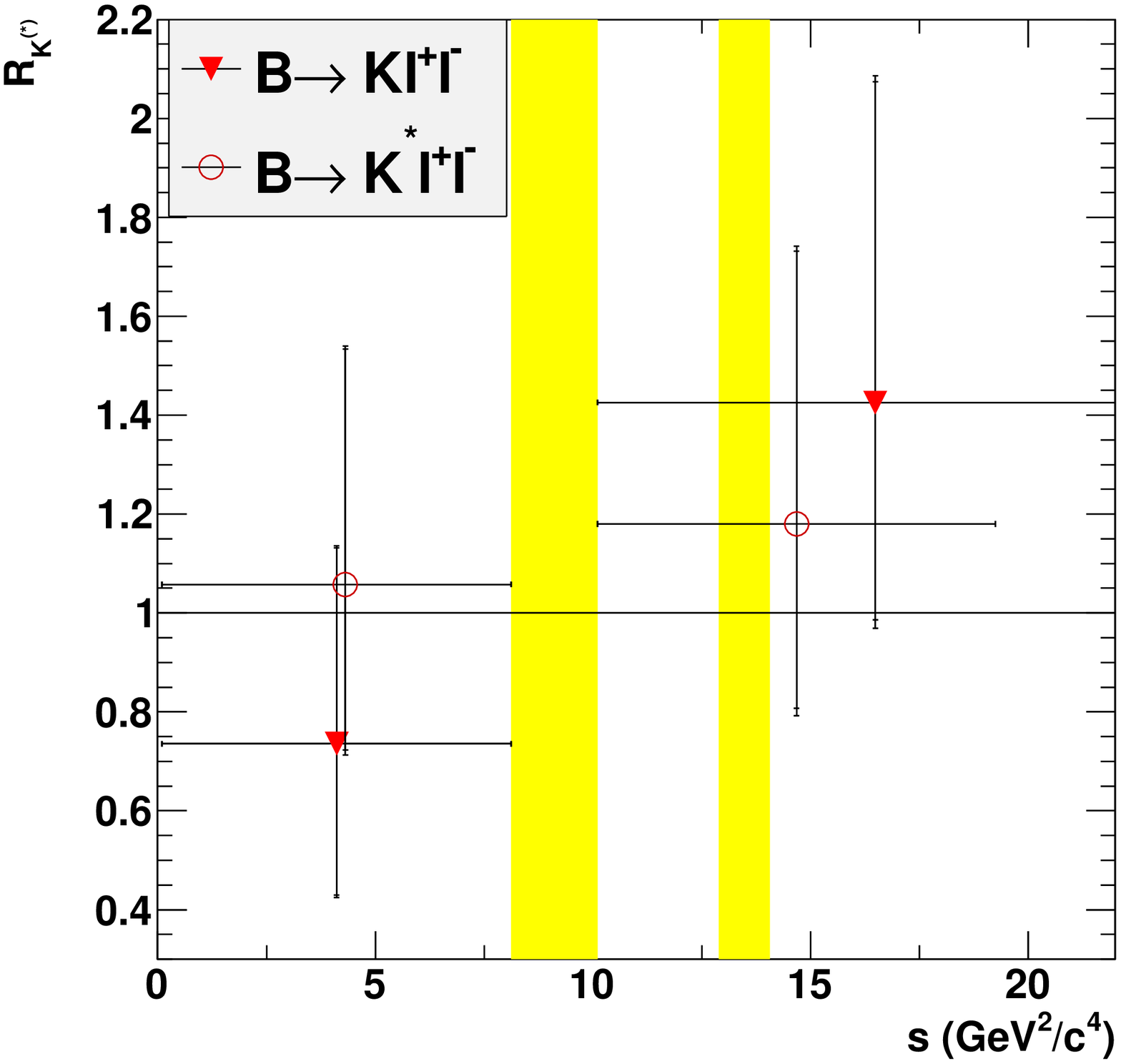}
\caption{\babar: (top) \CP asymmetries ${\cal A}_{CP}$ and (bottom) $R_{\Kmaybestar}$ in two $s$ regions for $K \ell^+ \ell^-$ and $K^* \ell^+ \ell^-$. The vertical 
       yellow shaded bands show the vetoed $s$ regions around the $J/\psi$ and $\psitwos$.~\cite{kllpaper}  }
\label{fig:acprk}
\end{center}
\end{figure}

In the SM, the direct $\CP$ asymmetry
\begin{eqnarray}
{\cal A}_{\CP}^{\Kmaybestar} \equiv
\frac
{{\cal B}(\overline{B} \rightarrow \overline{K}^{(*)}\ellell) - {\cal B}(B \rightarrow K^{(*)}\ellell)}
{{\cal B}(\overline{B} \rightarrow \overline{K}^{(*)}\ellell) + {\cal B}(B \rightarrow K^{(*)}\ellell)}
\end{eqnarray}
and lepton flavor ratio
\begin{eqnarray}
{\cal R}_{\Kmaybestar} \equiv
\frac
{{\cal B}(\kmaybemm)}
{{\cal B}(\kmaybeee)}
\end{eqnarray}
are expected to be very close to zero and one, respectively. Our measured
${\cal A}_{\CP}$ and ${\cal R}_{\Kmaybestar}$ results agree with the SM expectations,
    as shown in Fig~\ref{fig:acprk}.

\begin{figure}%[b!]
\begin{center}
\includegraphics[height=6cm]{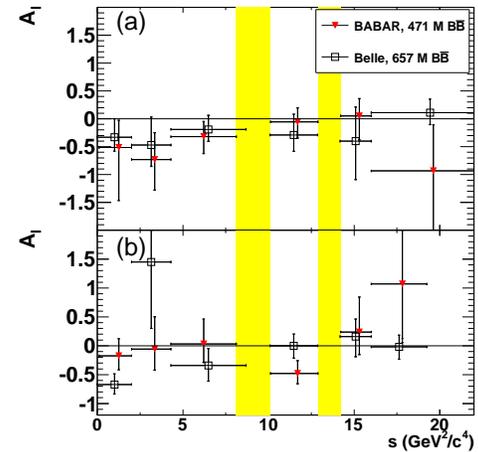}
\caption{\babar: Isospin asymmetry ${\cal A}_I$ for the (a) $K \ell^+ \ell^-$ and (b) $K^* \ell^+ \ell^-$ modes as a function of $s$, 
    in comparison to results from Belle~\protect\cite{belle09}. The vertical 
       yellow shaded bands show the vetoed $s$ regions around the $J/\psi$ and $\psitwos$ resonance.~\cite{kllpaper} }
\label{fig:isospin}
\end{center}
\end{figure}

The $\CP$-averaged isospin asymmetry is defined as:
\begin{equation}
%\resizebox{0.95\linewidth}{!}{$
{\cal A}^{K^{(*)}}_{I} \equiv
\frac
{{\cal B}(\Bz \to K^{(*)0}\ellell) - r_\tau {\cal B}(\Bp \to K^{(*)+}\ellell)} 
{{\cal B}(\Bz \to K^{(*)0}\ellell) + r_\tau {\cal B}(\Bp \to K^{(*)+}\ellell)},
%$}
\label{eq:aidef}
\end{equation}
\noindent
where $r_\tau \equiv \tau_{\Bz}/\tau_{\Bp}=1/(1.071\pm 0.009)$
is the ratio of $B^0$ and $B^+$ lifetimes~\cite{PDG}. 
In the SM, the isospin asymmetries are expected to be at the level of a few percent~\cite{isospin}. 
Figure~\ref{fig:isospin} shows the \babar\, isospin asymmetry results in agreement with
the earlier Belle results~\cite{belle09}. The corresponding LHCb results are presented in a separate FPCP 2012 talk~\cite{Serra:2012mb}.
This \babar\, analysis also measure below the $\jpsi$ resonance ($0.1<s<8.12$~\gevcccc):
\begin{eqnarray}
{\cal A}^{\rm low}_I(\modekavgll) &=&  -0.58_{-0.37}^{+0.29}\pm0.02,\nonumber\\
{\cal A}^{\rm low}_I(\modekstll)& = & -0.25_{-0.17}^{+0.20}\pm0.03,\nonumber
\end{eqnarray}
where the first and second uncertainties are statistical and systematic, respectively.
The above two results are consistent with the SM predictions at 2.1$\sigma$ and 1.2$\sigma$
significance levels, respectively. They also agree with the corresponding Belle results~\cite{belle09}.

\subsection{Angular Asymmetries}
\label{sec:kstllang}

For the angular observables in $\modekstll$,
   two angles are relevant here: 
  $\theta_K$ as the angle between the $K$ and the $B$ in
  the $K^*$ rest frame, and $\theta_{\ell}$ as the angle 
  between the $\ell^+$ ($\ell^-$) and the $B$ ($\Bbar$)
  in the $\ellell$ rest frame. The fraction
  of the $K^*$ longitudinal $\fl$ and the lepton forward-backward
  asymmetry $\afb$ are related to the distributions of $\ctk$ and $\ctl$
  in the signal $\modekstll$ decays through:
\begin{align*}
\frac{1}{\Gamma}\frac{d\Gamma}{d\cos\theta_{K}}&=\frac{3}{2}F_{L}\cos^{2}\theta_{K}+\frac{3}{4}\left(1-F_{L}\right)\left(1-\cos^{2}\theta_{K}\right),\\
%\resizebox{0.95\linewidth}{!}{$
\begin{split}    
\frac{1}{\Gamma}\frac{d\Gamma}{d\ctl}&=\frac{3}{4}F_{L}\left(1-\cos^{2}\theta_{\ell}\right)+\newln  \frac{3}{8}\left(1-F_{L}\right)\cdot\left(1  +\cos^{2}\theta_{\ell}\right)+\afb\cos\theta_{\ell}. 
\end{split}    
%$}
\end{align*}
The values of $\fl$ and $\afb$ can thus be extracted through the fits to these angular distributions.
Both $\fl$ and $\afb$ are well predicted in the SM, and follow distinct patterns as functions
of $s$~\cite{Ali:2002jg,ffmodels,flafbsmpred}. Starting from $s\rightarrow 0$,
   where $C_7^{\rm eff}$ dominates, the SM value of $\afb$ is negative at very low $s$, then cross the zero axis at $s\sim 4$\gevcccc. 
   At large $s$ above the \jpsi resonance, the SM value of $\afb$ is expected to be large and positive due to the domination of the product of $C_9^{\rm eff}$ and $C_{10}^{\rm eff}$.
   $\fl$ is only sensitive to the sign of $C_7^{\rm eff}$, 
   while $\afb$ is sensitive to both the sign of $C_7^{\rm eff}$ and that of $C_9^{\rm eff}\cdot C_{10}^{\rm eff}$.

\begin{figure}
\includegraphics[width=0.8\linewidth]{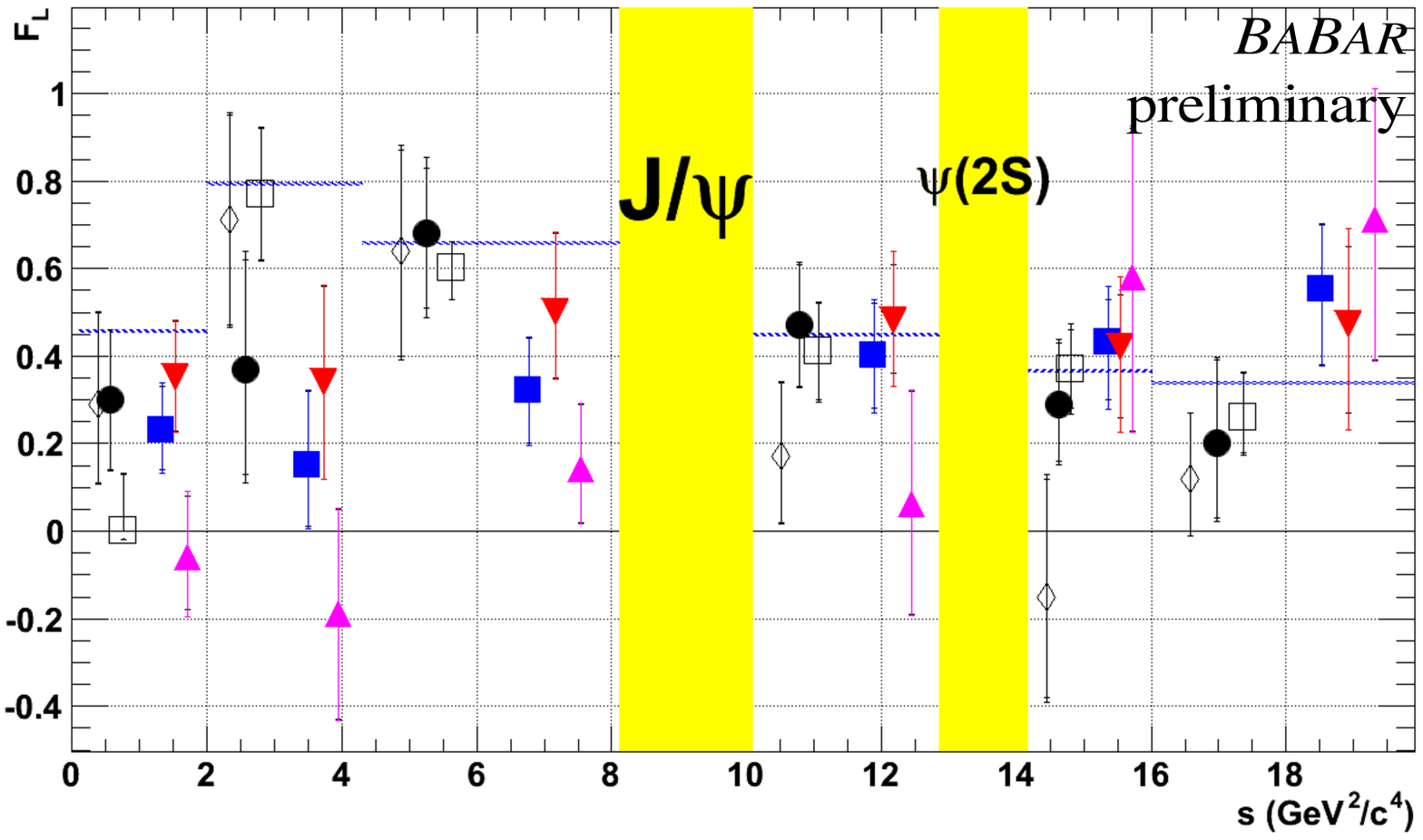}\\
\includegraphics[width=0.8\linewidth]{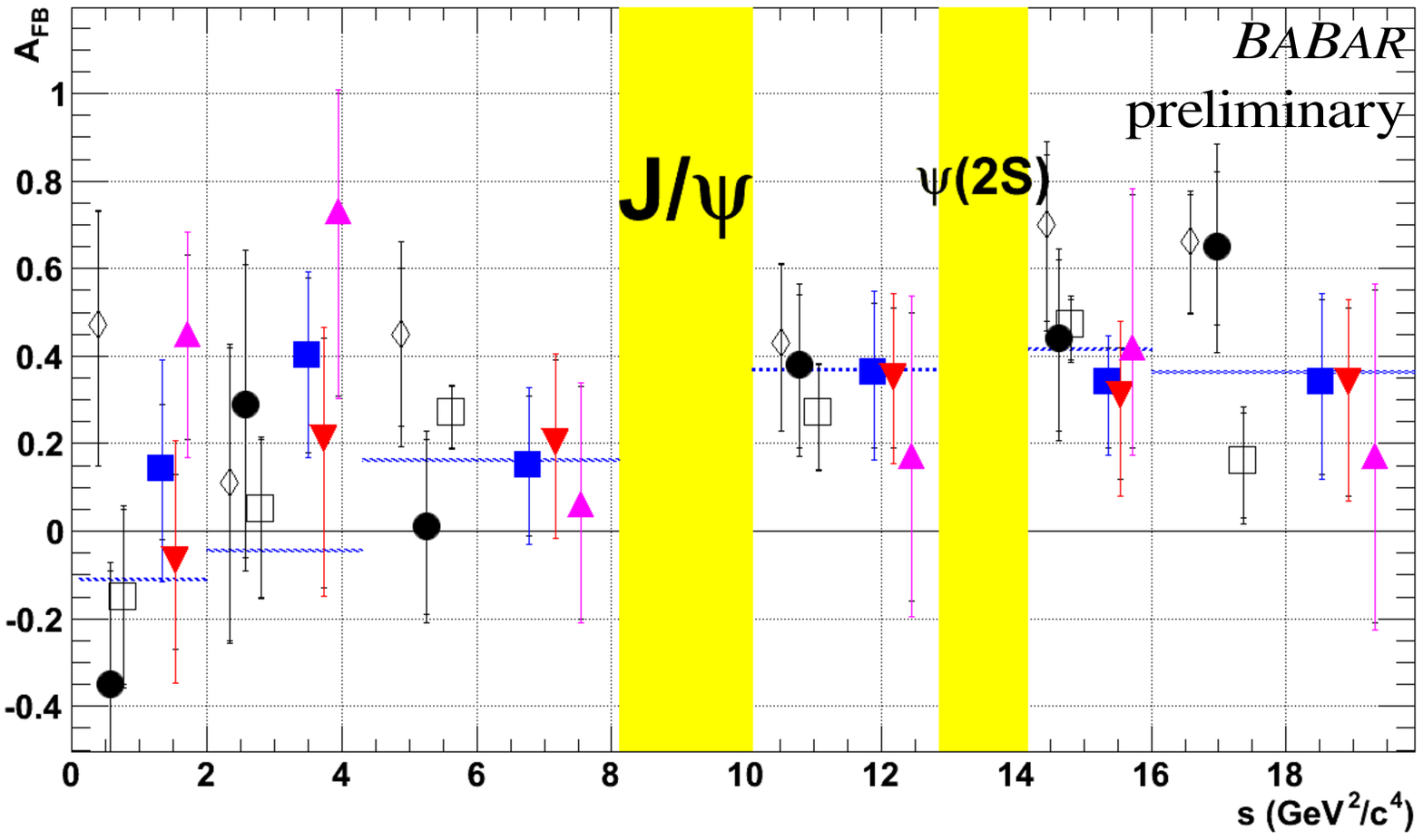}
\caption{\label{fig:flafb}\babar: Summary on (top) $\fl$ and (bottom) $\afb$ as a function of $s$ for $B\to \Kstar\ellell$.
Our results for $B^+\to\Kstarp\ellell$ (\textcolor{magenta}{$\blacktriangle$}), 
    $\Bz\to \Kstarz\ellell$ (\textcolor{red}{$\blacktriangledown$}),
   and all $B\to\Kstar\ellell$ modes combined ($\textcolor{blue}{\blacksquare}$)
   are compared to the results from Belle ($\lozenge$), CDF ($\bullet$), and LHCb ($\square$), as well as the SM predictions (blue dashed lines).
}
\end{figure}

   Figure~\ref{fig:flafb} shows the \babar\, $\fl$ and $\afb$ results for the $B^+\to\Kstarp\ellell$ modes, $\Bz\to \Kstarz\ellell$ modes,
   and all $B\to\Kstar\ellell$ modes combined, in the six $s$ bins. The $\fl$ and $\afb$ results for the combined $B\to\Kstar\ellell$
   are consistent with the most recent results from Belle~\cite{belle09}, CDF~\cite{cdf12afb}, and LHCb~\cite{lhcb11}, 
   as well as the SM predictions~\cite{Ali:2002jg,ffmodels,flafbsmpred}. Still some discrepancies
   with the SM are noticeable in the low $s$ bins below $\jpsi$ for the $B^+\to\Kstarp\ellell$ modes.

\section{Search for Lepton-Number Violating processes}
\begin{figure}
\includegraphics[width=0.8\linewidth]{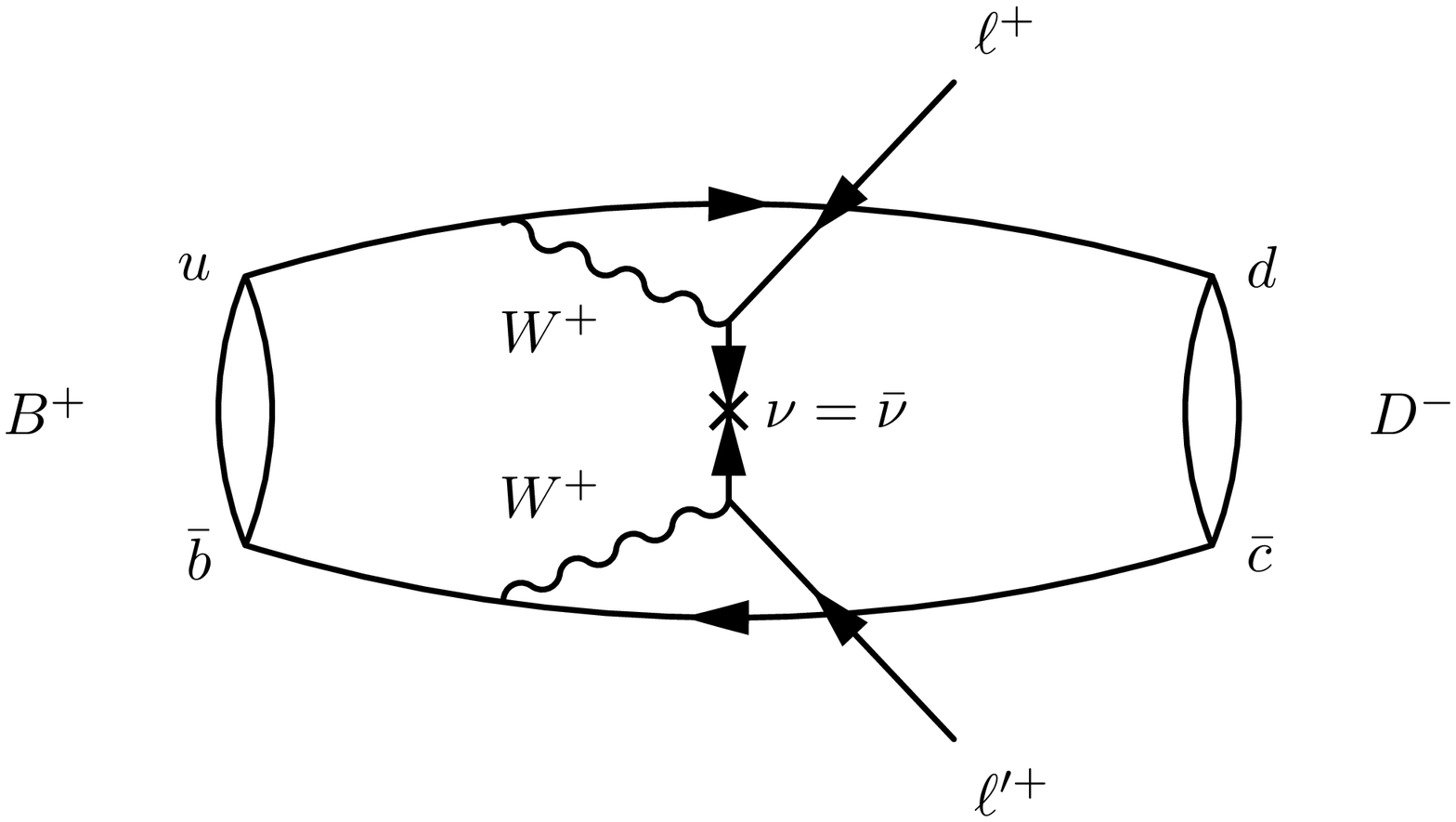}\\
\includegraphics[width=0.7\linewidth]{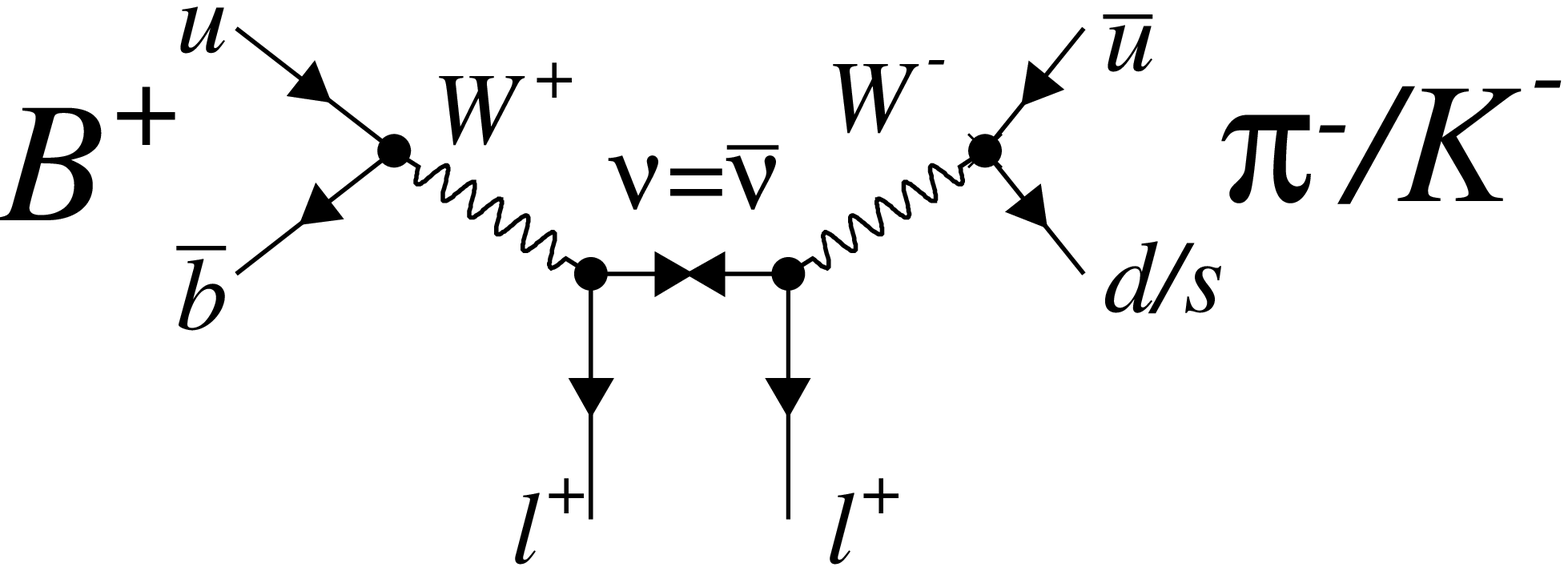}
\caption{\label{fig:lnvdiag}Possible Feynman diagrams for the decays (top) $\Bp \to D^- \ell^+\ell^+$~\cite{belledll}
    and (bottom) $\Bp \to (\pim/K^-) \ell^+\ell^+$~\cite{babarhll}.}
\end{figure}

In the SM with massless neutrinos, it is expected 
that the lepton number $L$ is conserved in low-energy collisions and decays. 
However, the observation of neutrino oscillation indicates that neutrinos possess non-zero
masses. If the neutrino masses are of Majorana type, \ie\, each neutrino is its own anti-particle,
the $L$ violation will be possible. Figure~\ref{fig:lnvdiag} shows two different types
of $\Delta L=2$ processes involving Majorana neutrinos: $\Bp \to D^- \ell^+\ell^+$,
   and $\Bp \to (\pim/K^-) \ell^+\ell^+$. The search on these decays is
an alternative approach to the search for neutrinoless double-beta decay.

\subsection{Search for $\Bp \to D^- \ell^+\ell^+$}
\begin{figure}
\includegraphics[width=0.8\linewidth]{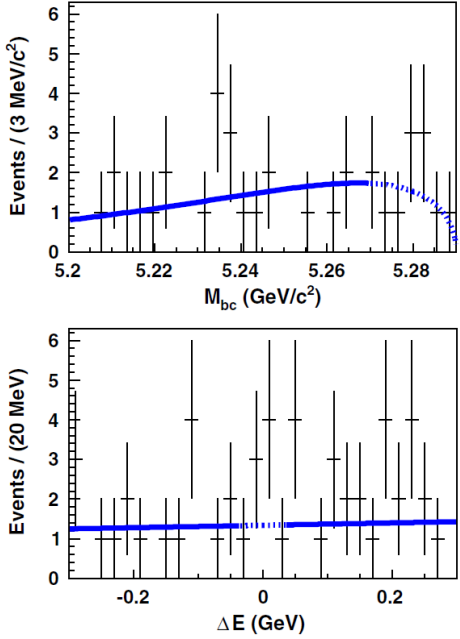}
\caption{\label{fig:dllfits}Belle: The 2D fit projections and data for (top) $M_{\rm bc}$ and (bottom) $\DeltaE$ in the $D^-\mu^+ \mu^+$ mode. 
    See Ref.\cite{belledll} for
        detailed description.}
\end{figure}

In this Belle analysis on $\Bp \to D^- \ell^+\ell^+$~\cite{belledll}, 
   $D^-$ meson is reconstructed through $D^-\to K^+\pi^-\pi^-$,
   and $\ell^+\ell^+$ is either $e^+ e^+$, $\mu^+ \mu^+$, or $e^+\mu^+$.
   Two kinematic variables are defined to identify signal $B$ decays: 
   the energy difference $\Delta E = E^*_B - E^*_{\rm beam}$ between
   the beam energy and the $B$ meson energy in the CM frame, and
   the beam-energy-constrained $B$ meson mass $M_{\rm bc}$, which
   is identical to $\mes$ used in the \babar\, experiment.
A two-dimensional (2D) fit is performed to the $M_{\rm bc}$-$\DeltaE$ distribution
for signal extraction. Figure~\ref{fig:dllfits} shows the fit projections
in the $D^-\mu^+ \mu^+$ mode. The 90\% C.L. upper limits on the branching fractions for different $\Bp \to D^- \ell^+\ell^+$
modes are $(1.0 - 2.6)\times 10^{-6}$ as summarized in Table~\ref{tab:dllUL}.

\begin{table}%[hbt]
  \caption{The 90\% C.L. upper
    limit (U.L.) on the branching fraction of $\Bp \to D^- \ell^+\ell^+$. 
        Here $\epsilon$ is the signal reconstruction efficiency, $N_{\rm obs}$ is the
    number of events in the signal region, $N_{\rm exp}^{\rm bkg}$ 
    is the expected number of background events in 
    the signal region.
    The efficiencies shown in the table do not include the 
    branching fraction of the $D^-$ decay.~\cite{belledll}
    }
  \label{tab:dllUL} 
%  \begin{ruledtabular}
    \begin{tabular}{ccccc}
    \hline\hline
      Mode & $\epsilon$ [\%] & $N_{\rm obs}$ & $N^{\rm bkg}_{\rm exp}$
      & U.L. [$10^{-6}$]\\\hline 
      $B^+\to D^- e^+e^+$ & 1.2 & 0 & 0.18$\pm$0.13 & $<$ 2.6 \\
      $B^+\to D^-e^+ \mu^+$ & 1.3 & 0 & 0.83$\pm$0.29 & $<$ 1.8 \\
      $B^+\to D^- \mu^+\mu^+$ & 1.9 & 0 & 1.44$\pm$0.43 & $<$ 1.0 \\
    \hline\hline
    \end{tabular}
%  \end{ruledtabular}
\end{table}

\subsection{Search for $\Bp \to h^- \ell^+\ell^+$}
\begin{figure}%[b]
\begin{center}
\includegraphics[width=0.8\linewidth]{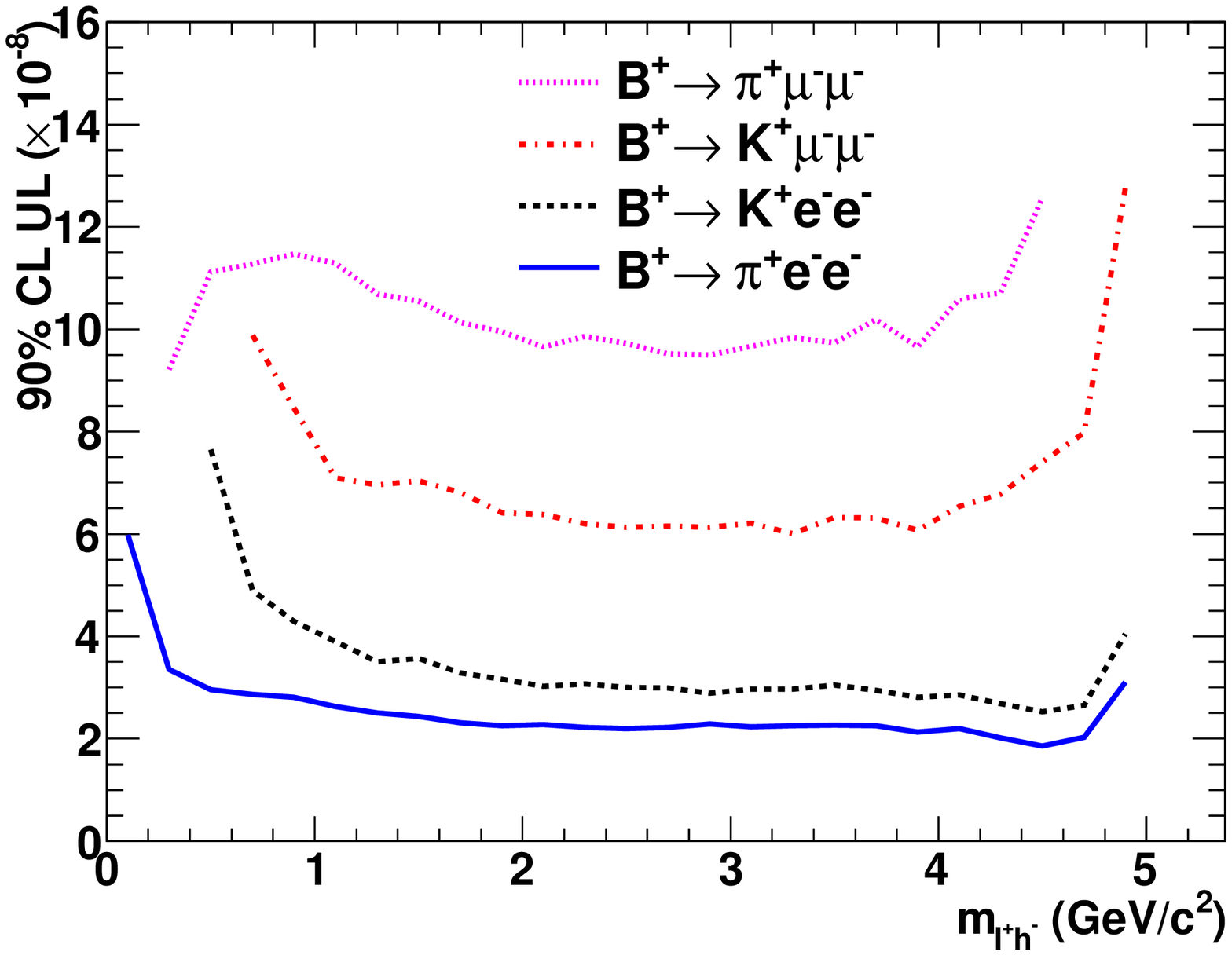}
\caption{\babar: 90\% C.L. upper limits on the branching fraction as a function
 of $m_{\ellp h^-}$ for $\Bp\to\pim\ep\ep$
 (blue solid line),  $\Bp\to\Km\ep\ep$  (black dashed line), $\Bp\to\pim\mup\mup$ 
 (red dash-dotted line), and  $\Bp\to\Km\mup\mup$ (magenta dotted line).~\cite{babarhll}
}
\label{fig:hllULscan}
\end{center}
\end{figure}

In this \babar\, analysis~\cite{babarhll}, the search is performed
in four different final states: $\pi^-e^+e^+$, $K^-e^+e^+$, $\pi^-\mu^+\mu^+$, and $K^-\mu^+\mu^+$.
Since these final states are very similar to the ones in the $B\to \Kmaybestar\ellell$ analysis described in
the previous section, both analyses also share similar event selection techniques. As no
significant evidence for signal events could be found, 
     this analysis sets the 90\% C.L. upper limits on the branching fractions for the four modes
as summarized in Table~\ref{tab:hllUL}. These results are comparable to the LHCb results~\cite{lhcblnv}
and provide much more stringent constraints than the CLEO results~\cite{cleolhv}.
The upper limits on the electron channels are also the most stringent ones to date.
Furthermore, Figure~\ref{fig:hllULscan} shows the 90\% C.L. upper limit on the branching fraction 
as a function of the invariant mass $m_{\ell^+h^-}$. If an exchange of the Majorana neutrino
occurs in the decay $\Bp\to h^- \ellp\ellp$ as illustrated in the left diagram of Fig.~\ref{fig:lnvdiag}, 
$m_{\ell^+h^-}$ could be interpreted as the Majorana neutrino mass.

\begin{table}%[htbp!]
\centering
\caption{ 90\% C.L. upper limits on the branching fractions (${\cal B}_{UL}$) for the four measured \B\ decays.~\cite{babarhll}}
\label{tab:hllUL}
\begin{tabular}{lrr}
\hline \hline
Mode & ${\cal B} (\times 10^{-8})$ & ${\cal B}_{UL} (\times 10^{-8})$ \\
\hline
$\Bp\to\pim\ep\ep$  & $0.27^{+1.1}_{-1.2}\pm 0.1$ & 2.3 \\
$\Bp\to\Km\ep\ep$  & $0.49^{+1.3}_{-0.8}\pm 0.1$ & 3.0 \\
$\Bp\to\pim\mup\mup$  & $0.03^{+5.1}_{-3.2}\pm 0.6$ & 10.7 \\
$\Bp\to\Km\mup\mup$  & $0.45^{+3.2}_{-2.7}\pm 0.4$ & 6.7 \\
\hline \hline
\end{tabular}
\end{table}

\section{Conclusion}
Both the \babar\, and Belle experiments are still actively producing interesting
results based on their full data sets. As $B$ factories, they are capable of inclusive analyses on the
$b\to s\gamma$ and $b\to s\ellell$ processes.
This report covers new results on rare $B$ decay channels 
from \babar\, and Belle. Generally all these
results are found to be consistent with the SM predictions,   
though some discrepancies exist for the angular asymmetries in the channel $B\to K^*\ellell$.
The current $B$ factory results are limited by low statistics, therefore
    we have to rely on LHCb and future B factories to continue new physics searches
    in these decay channels.

\bigskip % extra skip inserted
\begin{acknowledgments}
The author is supported by the United States National Science Foundation
grants.
Special thanks to the conference organizers for
their great work.
\end{acknowledgments}

\bigskip % extra skip inserted

\begin{thebibliography}{99}   % Use for  1-9  references
%\begin{thebibliography}{99} % Use for 10-99 references

%\bibitem{accelconf-ref}
%http://www.cern.ch/accelconf

%\bibitem{exampl-ref}
%A.N. Other, ``A Very Interesting Paper'', EPAC'96, Sitges, June
%1996.

%\bibitem{templates-ref}
%http://www.cern.ch/accelconf/templates.html
\bibitem{Misiak:2006zs}
M.~Misiak, H.~M.~Asatrian, K.~Bieri, M.~Czakon, A.~Czarnecki, T.~Ewerth, A.~Ferroglia and P.~Gambino {\it et al.},
  %``Estimate of B(anti-B ---> X(s) gamma) at O(alpha(s)**2),''
  Phys.\ Rev.\ Lett.\  {\bf 98}, 022002 (2007).
%  [hep-ph/0609232].

\bibitem{HFAG}
D.~Asner {\it et al.}  [Heavy Flavor Averaging Group Collaboration],
  %``Averages of b-hadron, c-hadron, and $\tau-lepton Properties,''
  arXiv:1010.1589 [hep-ex].

\bibitem{btosgbabar} 
  J.~P.~Lees {\it et al.}  [\babar\, Collaboration],
  %``Exclusive Measurements of $b \ra s \gamma$ Transition Rate and Photon Energy Spectrum,''
  arXiv:1207.2520 [hep-ex].

\bibitem{Benson:2004sg} 
  D.~Benson, I.~I.~Bigi and N.~Uraltsev,
  %``On the photon energy moments and their `bias' corrections in B ---> X(s) + gamma,''
  Nucl.\ Phys.\ B {\bf 710}, 371 (2005).

\bibitem{Lange:2005yw} 
  B.~O.~Lange, M.~Neubert and G.~Paz,
  %``Theory of charmless inclusive B decays and the extraction of V(ub),''
  Phys.\ Rev.\ D {\bf 72}, 073006 (2005).

\bibitem{Buchalla}
  G.~Buchalla, A.~J.~Buras, and M.~E.~Lautenbacher,
  %``Weak Decays Beyond Leading Logarithms,''
  Rev.\ Mod.\ Phys.\  {\bf 68}, 1125 (1996).
  %[arXiv:hep-ph/9512380].
  %%CITATION = RMPHA,68,1125;%%

\bibitem{isospin}
T.~Feldmann and J.~Matias,
  %``Forward-backward and isospin asymmetry for B --> K* l+ l- decay in the
  %standard model and in supersymmetry,''
  JHEP {\bf 0301}, 074 (2003).

\bibitem{belle09}
J.-T.~Wei {\it et al.}  [Belle Collaboration],
%``Measurement of the Differential Branching Fraction and Forward-Backword Asymmetry for B->K(*)l+l-''
Phys. Rev. Lett. {\bf 103}, 171801 (2009).

\bibitem{cdf11}
T.~Aaltonen {\it et al.}  [CDF Collaboration],
%``Observation of the Baryonic Flavor-Changing Neutral Current Decay Lambda_b
  %-> Lambda mu+ mu-,''
  Phys.\ Rev.\ Lett.\  {\bf 107}, 201802 (2011).

\bibitem{lhcb11}
R.~Aaij {\it et al.}  [LHCb Collaboration],
Phys.\ Rev.\ Lett.\  {\bf 108}, 181806 (2012).

\bibitem{kllpaper}
J.~P.~Lees {\it et al.}  [\babar\, Collaboration],
Phys.\ Rev.\ D {\bf 86}, 032012 (2012).

\bibitem{Ali:2002jg}
  A.~Ali, E.~Lunghi, C.~Greub, and G.~Hiller,
  %``Improved model-independent analysis of semileptonic and radiative rare  B
  %decays,''
  Phys.\ Rev.\  D {\bf 66}, 034002 (2002).

\bibitem{Zhong:2002nu}
M.~Zhong, Y.-L.~Wu, and W.-Y.~Wang,
Int. J. Mod. Phys. A {\bf 18}, 1959 (2003).  

\bibitem{ffmodels}
P.~Ball and R.~Zwicky,
%``New results on $B\to \pi, K, \eta$ decay form factors from light-cone sum rules''
Phys. Rev. D {\bf 71}, 014015 (2005);
%B.~Patricia and Z.~Roman,
%``New results on $B\to \pi, K, \eta$ decay form factors from light-cone sum rules''
Phys. Rev. D {\bf 71}, 014029 (2005).

\bibitem{PDG}
K. Nakamura {\it et al.}  [Particle Data Group],
  %``Review of particle physics,''
  J.\ Phys.\ G {\bf 37} (2010) 075021.

\bibitem{Serra:2012mb} 
  N.~Serra,
  %``Search for new physics in $B_s \rightarrow \mu^+ \mu^-$ and $B \rightarrow K^{(*)} \mu^+ \mu^-$,''
  arXiv:1208.3987 [hep-ex].

\bibitem{flafbsmpred}
A.~Ali, P.~Ball, L.~T.~Handoko and G.~Hiller,
  %``A Comparative study of the decays $B \to$ ($K$, $K^{*)} \ell^+ \ell^-$ in standard model and supersymmetric theories,''
  Phys.\ Rev.\ D {\bf 61}, 074024 (2000);
G.~Buchalla, G.~Hiller and G.~Isidori,
  %``Phenomenology of nonstandard $Z$ couplings in exclusive semileptonic $b \to s$ transitions,''
  Phys.\ Rev.\ D {\bf 63}, 014015 (2000);
F.~Kruger, L.~M.~Sehgal, N.~Sinha and R.~Sinha,
  %``Angular distribution and CP asymmetries in the decays anti-B ---> K- pi+ e- e+ and anti-B ---> pi- pi+ e- e+,''
 Phys.\ Rev.\ D {\bf 61}, 114028 (2000)
  [Erratum-ibid.\ D {\bf 63}, 019901 (2001)];
 F.~Kruger and J.~Matias,
  %``Probing new physics via the transverse amplitudes of B0 ---> K*0 (---> K- pi+) l+l- at large recoil,''
  Phys.\ Rev.\ D {\bf 71}, 094009 (2005).

\bibitem{cdf12afb}
 T.~Aaltonen {\it et al.}  [CDF Collaboration],
  %``Measurements of the Angular Distributions in the Decays $B \to K^{(*)} \mu^+ \mu^-$ at CDF,''
  Phys.\ Rev.\ Lett.\  {\bf 108}, 081807 (2012).

\bibitem{belledll}  
O.~Seon, Y.~J.~Kwon, T.~Iijima, I.~Adachi, H.~Aihara, D.~M.~Asner, T.~Aushev and A.~M.~Bakich {\it et al.}  [Belle Collaboration],
  %``Search for Lepton-number-violating B+->D-l+l'+ Decays,''
  Phys.\ Rev.\ D {\bf 84}, 071106 (2011).

\bibitem{babarhll}
 J.~P.~Lees {\it et al.}  [\babar\, Collaboration],
  %``Search for lepton-number violating processes in B+ -> h- l+ l+ decays,''
  Phys.\ Rev.\ D {\bf 85}, 071103 (2012).

\bibitem{lhcblnv}
R.~Aaij {\it et al.}  [LHCb Collaboration],
  %``Search for the lepton number violating decays $B^{+}\to \pi^- \mu^+ \mu^+$ and $B^{+}\to K^- \mu^+ \mu^+$,''
  Phys.\ Rev.\ Lett.\  {\bf 108}, 101601 (2012).

\bibitem{cleolhv}
K.~W.~Edwards {\it et al.}  [CLEO Collaboration],
  %``Search for lepton flavor violating decays of $B$ mesons,''
  Phys.\ Rev.\ D {\bf 65}, 111102 (2002).

\end{thebibliography}

\end{document}